\documentstyle[manuscript,aps]{revtex}
\begin{document}


\title{Semiclassical Analysis of Quasi-Exact Solvability}

\author{Carl M. Bender}
\address{Department of Physics, Washington University, St. Louis, MO 63130,
USA}

\author{Gerald V. Dunne}
\address{Physics Department, University of Connecticut, Storrs, CT 06269, USA}

\author{Moshe Moshe}
\address{Department of Physics, Technion -- Israel Institute of Technology,
Haifa 32000, Israel}

\maketitle

\begin{abstract}
Higher-order WKB methods are used to investigate the border between the
solvable
and insolvable portions of the spectrum of quasi-exactly solvable
quantum-mechanical potentials. The analysis reveals scaling and factorization
properties that are central to quasi-exact solvability. These two properties
define a new class of semiclassically quasi-exactly solvable potentials.
\end{abstract}

Quantum-mechanical potentials are said to be {\em quasi-exactly solvable} (QES)
if it is possible to find a finite portion of the energy spectrum and
associated
eigenfunctions exactly and in closed form \cite{Ush}. QES potentials depend on
a
parameter $J$; for positive integer values of $J$ one can find exactly the
first
$J$ eigenvalues and eigenfunctions, typically of a given parity. QES systems
have been classified using an algebraic approach in which the Hamiltonian is
expressed in terms of the generators of a Lie algebra \cite{Tur,ST,GKO}. This
approach generalizes the dynamical symmetry analysis of {\em exactly solvable}
quantum-mechanical systems, whose {\em entire} spectrum may be found in closed
form by algebraic means \cite{Iac}.

In this paper we use higher-order semiclassical methods to examine the boundary
between the exactly solvable part of the spectrum and the remaining energy
levels in QES systems. We find that the large-$J$ asymptotic behavior of the
largest exactly known energy eigenvalue is particularly simple for QES
potentials. This study leads to a natural generalization of QES potentials. We
discover an infinite tower of potentials; the first is exactly solvable, the
second is QES, and the rest share some semiclassical features of QES potentials
but are not QES.

The simplest QES potential \cite{BD} is
\begin{eqnarray}
V(x)=x^6-(4J-1)x^2.
\label{e1}
\end{eqnarray}
The Schr\"odinger equation, $-\psi''(x)+[V(x)-E]\psi(x)=0$,
has $J$ even-parity solutions of the form
\begin{eqnarray}
\psi(x)=e^{-x^4/4}\sum_{k=0}^{J-1}c_k x^{2k}.
\label{e2}
\end{eqnarray}
The coefficients $c_k$ satisfy the recursion relation
\begin{eqnarray}
4(J-k)c_{k-1}+Ec_k+2(k+1)(2k+1)c_{k+1}=0,
\label{e3}
\end{eqnarray}
where $0\leq k\leq J-1$ and we define $c_{-1}=c_{J}=0$.

The simultaneous linear equations (\ref{e3}) have a nontrivial solution for
$c_0,\,c_1,\,...,\,c_{J-1}$ if the determinant of the coefficients vanishes.
For
each integer $J$ this determinant is a polynomial $P_J(E)$ of degree $J$ in the
variable $E$. The roots of $P_J(E)$ are all real and are the $J$ quasi-exact
energy eigenvalues of the potential in Eq.~(\ref{e1}). We have computed the
first 40 polynomials; we list the first seven of these below:
\begin{eqnarray}
P_1(E) &=& E,\nonumber\\
P_2(E) &=& -8+E^2,\nonumber\\
P_3(E) &=& -64E+E^3,\nonumber\\
P_4(E) &=& 2880-240E^2+E^4,\nonumber\\
P_5(E) &=& 47104E-640E^3+E^5,\nonumber\\
P_6(E) &=& -5184000+331456E^2-1400E^4+E^6,\nonumber\\
P_7(E) &=& -130940928E+1529856E^3-2688E^5+E^7.
\nonumber
\end{eqnarray}

The roots of $P_J(E)$ occur in positive and negative pairs, and successive sets
of roots interlace:
\begin{eqnarray}
{\rm roots~of}~P_1(E) &:& \,\, 0;\nonumber\\
{\rm roots~of}~P_2(E) &:& \,\, \pm 2\sqrt{2};\nonumber\\
{\rm roots~of}~P_3(E) &:& \,\, 0,~\pm 8;\nonumber\\
{\rm roots~of}~P_4(E) &:& \,\, \pm 3.55932,~\pm 15.07751;\nonumber\\
{\rm roots~of}~P_5(E) &:& \,\, 0,~\pm 9.21135,~ \pm 23.56164;\nonumber\\
{\rm roots~of}~P_6(E) &:& \,\, \pm 4.10132,~\pm 16.70778,~ \pm
33.22693;\nonumber\\
{\rm roots~of}~P_7(E) &:& \,\, 0,~ \pm 10.18750,~\pm 25.56258,~ \pm 43.94052.
\nonumber
\end{eqnarray}

Let $E_{\rm c}(J)$ represent the largest root and thus the {\em critical}
energy
that marks the upper edge of the quasi-exact spectrum. A numerical fit to the
large-$J$ asymptotic behavior of $E_{\rm c}(J)$ for $1\leq J\leq 40$ using
Richardson extrapolation \cite{BO8} gives $E_{\rm c}(J)\sim 3.0792014355 J^
{3/2}$ as $J\to\infty$. We recognize the numerical constant:
\begin{eqnarray}
E_{\rm c}(J)\sim {16\over 9}\sqrt{3}\, J^{3/2}\quad (J\to\infty).
\label{e6}
\end{eqnarray}

One may verify this result analytically by finding the minimum of the potential
$V(x)=x^6-(4J-1)x^2$. Minima occur at $x=\pm [(4J-1)/3]^{1/4}$; at these values
$V\sim -{16\over 9}\sqrt{3}J^{3/2}~(J\to\infty)$. Since the quasi-exact energy
eigenvalues occur in $\pm$ pairs and the zero-point energy is negligible for
large $J$, the asymptotic result in Eq.~(\ref{e6}) is confirmed \cite{XX}.
However, this approach is not useful for all QES systems because in general the
spectrum is not symmetric under $E\to-E$.

WKB theory provides a more general derivation of Eq.~(\ref{e6}) and gives
higher-order corrections. Furthermore, semiclassical analysis reveals features
of $V(x)$ that are generic to QES potentials. The leading-order WKB
quantization
condition is
\begin{eqnarray}
(n+\frac{1}{2})\pi\sim\int_{-x_0}^{x_0}dx\,\sqrt{E-x^6+(4J-1)x^2}
\label{e7}
\end{eqnarray}
for large $n$, where $\pm x_0$ are turning points [$x_0$ is the positive zero
of
$E-V(x)$] and $n$ is the quantum number of the energy level. We seek the $J$th
{\em even}-parity energy level $E=E_{\rm c}(J)$ for which $n=2J$.

To evaluate the integral on the right side of Eq.~(\ref{e7}) for large $J$, we
begin by scaling the variables $E_{\rm c}$ and $x$. The scaling $E_{\rm c}=
J^{3/2}\alpha$ and $x=J^{1/4}y$ extracts a factor of $J$ from the integral and
reduces Eq.~(\ref{e7}) to an exact expression for the numerical constant
$\alpha$:
\begin{eqnarray}
2\pi=\int_{-y_0}^{y_0}dy\,\sqrt{\alpha-y^6+4y^2},
\label{e8}
\end{eqnarray}
where $y_0$ is the positive root of $\alpha -y^6+4y^2=0$.

In general, the integral in  Eq.~(\ref{e8}) is not an elementary function of
$\alpha$. However, for the special value $\alpha={16\over 9}\sqrt{3}$ the
polynomial $\alpha-y^6+4y^2$ {\em factors into a product of a linear term times
a perfect square}:
\begin{eqnarray}
{16\over 9}\sqrt{3}-y^6+4y^2=\left({4\over\sqrt{3}}-y^2\right)\left(
{2\over\sqrt{3}}+y^2\right)^2.
\label{e9}
\end{eqnarray}
Using this factorization, we express the integral in Eq.~(\ref{e8}) in terms of
Beta functions:
\begin{eqnarray}
&& \int_{-y_0}^{y_0}dy\,\sqrt{\alpha-y^6+4y^2} = {8\over 3}\int_{0}^{1} dz\,
{1+2z\over\sqrt{z}}\sqrt{1-z}\nonumber\\
&& \qquad ={8\over 3}B\left({3\over 2},{1\over 2}\right)
+{16\over 3}B\left({3\over 2},{3\over 2}\right)=2\pi.
\label{e10}
\end{eqnarray}
Thus, the WKB quantization condition (\ref{e8}) is satisfied with $\alpha=
\frac{16}{9}\sqrt{3}$, which confirms the leading asymptotic form of the
critical energy $E_{\rm c}(J)$ given in Eq.~(\ref{e6}).

Two aspects of the above calculation of the WKB integral are crucial. First,
the
QES potential is such that in the limit $J\to\infty$, the parameter $J$ scales
out of the integrand. Second, the factorization in Eq.~(\ref{e9}) enables us to
evaluate the scaled integral and to obtain the factor of $\pi$, which then
cancels from the leading-order WKB quantization condition. In fact, merely
demanding that $\alpha -y^6+4y^2$ factor into the product of a linear term in
$y^2$ times a square of a linear term in $y^2$ {\em uniquely specifies} the
value of $\alpha$, which in turn determines the large-$J$ asymptotic behavior
of the critical energy $E_{\rm c}(J)$.

The factorization property extends to general QES $x^6$ potentials. Consider
the
scaled factored form $\alpha -V(y)=(a-y^2)(b+cy^2)^2$. (The conditions $a>0$,
$b,\,c\geq 0$ ensure that there is just one pair of turning points and that the
WKB integrals give rise to Beta functions of half-integer arguments.) The
leading-order WKB condition is
\begin{eqnarray}
2\pi J &\sim& \int dx\sqrt{E-V(x)}\nonumber\\
&=& J\int dy\sqrt{a-y^2}(b+cy^2)= 2\pi J({ab\over 4}+{a^2 c\over 16}),
\nonumber
\end{eqnarray}
which determines $b$ in terms of  $a$ and $c$: $b={4\over a}-{ac\over 4}$. This
gives the potential
$$V(y)=c^2y^6+2\beta c y^4+(\beta^2-4c)y^2,$$
where $\beta={4\over a}-{3ac\over 4}$. This is exactly the large-$J$ scaled
form
of the {\em general} nonsingular QES $x^6$ potential (class VI in
Ref.~\cite{Tur}). When $\beta=0$ and $c=1$, we obtain the case treated in
Eq.~(\ref{e1}).

This factorization is a universal feature of quasi-exact solvability. In
general, at the upper end of the quasi-exact spectrum, $E-V(x)$ factors into
one
term, which fixes the location of the two WKB turning points, multiplied by a
second term, which is a perfect square. As another example, consider the QES
potential $V(x)=\sinh^2 x -(2J-1)\cosh x$. For this potential there are $J$
even-parity QES states, all of the form $\psi(x)=e^{-\cosh x}P(\cosh x)$, where
$P$ is a polynomial. To find the critical energy $E_{\rm c}$ at the upper
boundary of the quasi-exact spectrum, we apply the WKB quantization condition
\begin{eqnarray}
\int_{-x_0}^{x_0}dx && \,\sqrt{E_{\rm c}-\sinh^2(x)+(2J-1)\cosh(x)}\nonumber\\
&&\qquad\qquad\qquad \sim (2J+1/2)\pi \quad (J\to\infty).
\label{e11}
\end{eqnarray}
The substitution $x=2t$ reveals the factorization property of the integrand in
Eq.~(\ref{e11}). The expression $E_{\rm c}-V(x)$ factors into one term $4J-2-4
\cosh^2 t$ (this term determines the turning points) multiplied by another term
in the form of a perfect square, $\sinh^2 t$. This factorization fixes the
dependence of $E_{\rm c}$ on $J$: $E_{\rm c}=1-2J$. With this factorization,
the
right side of Eq.~(\ref{e11}) simplifies to $4\int_0^{t_0}dt\,\sinh t\sqrt{4J-2
-4\cosh^2 t}\sim 2\pi J$, for large $J$. Once again, $\pi$ divides out of the
leading-order WKB quantization condition.

The remarkable connection between WKB and quasi-exact solvability persists to
higher order in WKB theory. The factor of $\pi$ cancels from the WKB
quantization condition {\em to all orders} leaving a purely algebraic series.
To
illustrate this, we do a fifth-order WKB calculation of $E_{\rm c}(J)$ for the
$x^6$ potential in Eq.~(\ref{e1}). We begin by scaling the energy in the
Schr\"odinger equation using $E_{\rm c}(J)=2[(4J-1)/3]^{3/2}\gamma$.
After scaling the independent variable, we obtain the Schr\"odinger equation
\begin{eqnarray}
\epsilon^2\psi''(u)=(4u^6-3u^2-\gamma)\psi(u),
\label{e12}
\end{eqnarray}
where the small parameter is $\epsilon=3/[2(4J-1)]$. The series representation
for the scaled energy $\gamma$ is
\begin{eqnarray}
\gamma=\sum_{n=0}^{\infty}a_n\epsilon^n,
\label{e13}
\end{eqnarray}
where $a_0=1$ as a result of the factorization in Eq.~(\ref{e9}).

The WKB quantization condition to order $\epsilon^5$ is \cite{BO}
\begin{eqnarray}
\left(2J+{1\over 2}\right) \pi &=& {1\over 2\epsilon}\oint
du\,Q^{1/2}-{\epsilon
\over 96}\oint du\,{Q''\over Q^{3/2}}\nonumber\\
&-&{\epsilon^3\over 3072}\oint du\,{2Q''''Q-7(Q'')^2\over Q^{7/2}}+\dots,
\label{e14}
\end{eqnarray}
where $Q(u)=\gamma+3u^2-4u^6$ and where the contours encircle the two turning
points (it is crucial that there be only two) and the branch cut joining them.

Next, we expand Eq.~(\ref{e14}) in powers of $\epsilon$ and perform the
resulting contour integrals. Each of these integrals can be evaluated in closed
form because of the leading-order factorization $f(u)\equiv -4u^6+3u^2
+1=(1-u^2)(2u^2+1)^2$. Indeed, all integrals of the form $\oint du\, u^{2m}
[f(u)]^{1/2-n}$ where $m$ and $n$ are nonnegative integers can be expressed as
simple multiples of $\pi$. For example,
\begin{eqnarray}
\oint du\,[f(u)]^{1/2} &=& {3\pi\over 2},\quad \oint du\,[f(u)]^{-1/2}=
-{2\pi\over\sqrt{3}},\nonumber\\
\oint du\,[f(u)]^{-3/2} &=& -{10\pi\over 9\sqrt{3}},\quad
\oint du\,u^4 [f(u)]^{-3/2} ={\pi\over 18\sqrt{3}}.
\nonumber
\end{eqnarray}

We evaluate the appropriate contour integrals and then use Eq.~(\ref{e14}) to
determine the first five expansion coefficients $a_n$ for the scaled energy
$\gamma$ in Eq.~(\ref{e13}). The expansion for the critical energy
$E_{\rm c}(J)$ is
\begin{eqnarray}
E_{\rm c}(J) &\sim& 2\left({4J-1\over 3}\right)^{3/2}\biggm[1-{3\sqrt{3}\over
4J
-1} \nonumber\\
&& +{35\over 8(4J-1)^2}+{5\sqrt{3}\over 3(4J-1)^3} \nonumber\\
&& +{23281\over 3456(4J-1)^4}+{88945\sqrt{3}\over 5184(4J-1)^5}+...\biggm].
\label{e17}
\end{eqnarray}

Numerical results are excellent; when $J=40$ the exact value of $E_{\rm c}(J)$
is $746.606715392...$, while the fifth-order WKB result in Eq.~(\ref{e17})
gives
$746.606715384...\,$.

The factorizability property of QES potentials has the consequence that
semiclassical analysis can be performed to all orders. All integrals give
simple
multiples of $\pi$ and no transcendental functions ever appear. This is in
stark
contrast with $x^{2N}$ anharmonic oscillator potentials for which WKB
analysis leads to elliptic functions \cite{HM}.

Although factorizability is a property of QES potentials, it extends to a
larger
class of potentials that are not QES. We call such potentials {\em
semiclassically quasi-exactly solvable} (SQES). Consider the potential
\begin{eqnarray}
V(x)=x^{10}-4Jx^4-\delta J^{4/3}x^2.
\label{e18}
\end{eqnarray}
For fixed $\delta$ this potential is {\em not} QES in the conventional
group-theoretic sense \cite{Tur,ST,GKO}; one cannot find more than one exact
algebraically determined eigenstate and corresponding eigenvalue. Nevertheless,
for large $J$ we can impose the scaling and factorizability conditions.
Generalizing from the factorization of the $x^6$ potential in Eq.~(\ref{e9}),
we
demand that the quantity $E-V(x)$ factor into a product of a linear term in
$x^2$ multiplied by a perfect square. This requirement fixes the numerical
value
of $\delta=3(4/5)^{1/3}$ and the asymptotic behavior of the critical energy
$E_{\rm c}(J)\sim{9\over 2}(4J/5)^{5/3}$ ($J\to\infty$), and we have
\begin{eqnarray}
&& {9\over 2}{4J\over 5}^{5\over 3}-x^{10}+4Jx^4+{15\over 4}\left({4J\over 5}
\right)^{4\over 3}x^2\nonumber\\
&& =J^{5\over 3}\left[2\left({4\over 5}\right)^{1\over
3}{\hskip-2pt}-y^2\right]
{\hskip-4pt}\left[y^4+\left({4\over 5}\right)^{1\over 3}{\hskip-2pt}y^2
+{3\over 2}\left({4\over 5}\right)^{2\over 3}\right]^2{\hskip-1pt},{\hskip-1pt}
\label{e19}
\end{eqnarray}
where $x=J^{1/6}y$. Because the squared factor has no real zeros, there are
{\em exactly} two real turning points whose locations are determined by the
linear factor in $y^2$.

This factorization enables us to evaluate the large-$J$ leading-order WKB
quantization integral exactly:
\begin{eqnarray}
2\pi J &\sim& \int_{-x_0}^{x_0}dx\,\sqrt{E-V(x)}\nonumber\\
&=&\left[{32\over 5}B\left({3\over 2},{5\over 2}\right)+{16\over 5}B\left({3
\over 2},{3\over 2}\right)+{20\over 5}B\left({3\over 2},{1\over
2}\right)\right]
J.
\nonumber
\end{eqnarray}
As in the case of the QES potential in Eq.~(\ref{e1}), the higher-order WKB
quantization integrals can be done exactly.

The potential in Eq.~(\ref{e18}) belongs to an infinite hierarchy of SQES
$x^{4k+2}$ potentials that exhibit factorization and scaling properties but
which are not QES for $k=2,~3,~4,~...\,$. The $k$th such potential is
\cite{SIM}
\begin{eqnarray}
&& V(x)={16(k+1)^2\over (2k+1)^2}\left[{\Gamma(k+3/2)\over 2\sqrt{\pi}
\Gamma(k+2)}\right]^{1\over k+1}J^{2k+1\over k+1}\Biggm\{1 \nonumber\\
&& -{\hskip-4pt}\left[{}_2F_1{\hskip-4pt}\left({1\over 2},{\hskip-2pt}
-{1\over 2}-k;{1\over 2}-k;{\hskip-4pt}
\left[{\Gamma(k+3/2)\over 2\sqrt{\pi}\Gamma(k+2)J}\right]^{1\over k+1}
{\hskip-6pt}x^2{\hskip-2pt}\right){\hskip-3pt}
\right]^2{\hskip-3pt}\Biggm\},
\nonumber
\end{eqnarray}
where ${}_2F_1$ is a hypergeometric function \cite{AS}.

To demonstrate the factorization property we substitute $y^2=
\left\{\Gamma(k+3/2)/[2\sqrt{\pi}\Gamma(k+2)J]\right\}^{1/(k+1)}x^2$
and express $E-V(x)$ as a linear factor in $y^2$ multiplied by a square of a
polynomial of degree $k$ in $y^2$:
\begin{eqnarray}
&&{16(k+1)^2\over (2k+1)^2}\left[{\Gamma(k+3/2)\over 2\sqrt{\pi}\Gamma(k+2)}
\right]^{1\over k+1} J^{2k+1\over k+1} -V(x)\nonumber\\
&& = \left[{2J\sqrt{\pi}\Gamma(k+2)\over\Gamma(k+3/2)}\right]^{2k+1\over k+1}
{\hskip-6pt}\left(1-y^2\right){\hskip-4pt}
\left[\sum_{n=0}^k {\Gamma(n+{1\over 2})\over n!\sqrt{\pi}}y^{2k-2n}\right]^2
{\hskip-3pt}.
\nonumber
\end{eqnarray}
Note that there are two real turning points at $y=\pm 1$.

This factorization fixes the large-$J$ asymptotic behavior of the critical
energy:
\begin{eqnarray}
E_{\rm c}\sim {16(k+1)^2\over (2k+1)^2}\left[{\Gamma(k+3/2)\over 2\sqrt{\pi}
\Gamma(k+2)}\right]^{1\over k+1}J^{2k+1\over k+1}.
\label{e22}
\end{eqnarray}
Leading-order WKB verifies this asymptotic behavior:
\begin{eqnarray}
2\pi J &\sim& \int_{-x_0}^{x_0}dx\,\sqrt{E-V(x)}\nonumber\\
&=& J{2\sqrt{\pi}(k+1)!\over\Gamma(k+3/2)} \int_0^1 du\, {\sqrt{1-u}\over
\sqrt{u}} \sum_{n=0}^k{\Gamma(n+1/2)\over n!\sqrt{\pi}}u^{k-n}\nonumber\\
&=&J{2\sqrt{\pi}(k+1)!\over\Gamma(k+3/2)}\sum_{n=0}^k{\Gamma(n+1/2)\over n!
\sqrt{\pi}}B\left({3\over 2},k-n+{1\over 2}\right).
\nonumber
\end{eqnarray}

Let us examine the hierarchy of SQES potentials for integer \cite{INT} values
of
$k$ using the general formula \cite{AS}
\begin{eqnarray}
&& {}_2F_1(1/2,-k-1/2;-k+1/2;z)\nonumber\\
&&\qquad = {\Gamma(1/2-k)\over\Gamma(1/2)}z^{k+1/2}\left({d\over dz}\right)^k
\left[ {(1-z)^{k+1/2}\over\sqrt{z}}\right].
\label{e23}
\end{eqnarray}
For $k=0$ we have $E_{\rm c}\sim 4J$ and ${}_2F_1\left({1\over 2},-{1\over
2};{1
\over 2};z\right)=\sqrt{1-z}$, and we obtain the {\em exactly solvable}
harmonic
oscillator potential, $V(x)=x^2$. For $k=1$ we have $E_{\rm c}\sim {16\over 9}
\sqrt{3}J^{3/2}$ and ${}_2F_1\left({1\over 2},-{3\over 2};-{1\over 2};z\right)=
(1+2z)\sqrt{1-z}$, and we obtain the large-$J$ asymptotic approximation to the
$x^6$ QES potential in Eq.~(\ref{e1}): $V(x)=x^6-4Jx^2$. For $k=2$ we have $E_
{\rm c}\sim{9\over 2}(4J/5)^{5/3}$ and ${}_2F_1\left({1\over 2},-{5\over 2};-{3
\over 2};z\right)=\left(1+{4\over 3}z+{8\over 3}z^2\right)\sqrt{1-z}$, and we
obtain the $x^{10}$ SQES potential in Eq.~(\ref{e19}): $V(x)=x^{10}-4Jx^4
-{15\over 4}(4J/5)^{4/3}x^2$. In the limit $k\to\infty$ we obtain the
square-well potential, for which $E_{\rm c}\sim 4 J^2$. Thus, at the bottom of
the SQES hierarchy is the exactly solvable harmonic oscillator, followed by the
QES $x^6$ potential, which is then followed by the range of new SQES potentials
and, finally, in the $k\to\infty$ limit, by the square-well potential.

For QES potentials the critical energy $E_{\rm c}(J)$ lies at the upper
boundary of the quasi-exact spectrum. For SQES potentials, the critical energy
$E_{\rm c}(J)$ is still a significant point in the energy spectrum; at this
point we observe a sharp change (a first-order phase transition for large $J$)
in the density of states. If we use ordinary WKB analysis to find the $2n$th
energy eigenvalue for a $x^{4k+2}$ potential, where $n$ is not correlated with
$J$, we find that as $n\to\infty$
\begin{eqnarray}
E_{2n}\sim\left[{4\sqrt{\pi}(k+1)n\Gamma\left({k+1\over 2k+1}\right)\big/
\Gamma\left({1\over 4k+2}\right)}\right]^{2k+1\over k+1}.
\nonumber
\end{eqnarray}
Observe that the numerical coefficient in this asymptotic relation is a
different function of $k$ from that in Eq.~(\ref{e22}). For example, for $k=1$
we have $E_{\rm c}\sim 3.0792J^{3/2}$ but $E_{2n}\sim 6.4066n^{3/2}$. Also, for
$k=2$ we have $E_{\rm c}\sim 3.1024J^{5/3}$ while $E_{2n}\sim 7.4235 n^{5/3}$.
As $k\to\infty$ we have $E_{\rm c}\sim 4 J^2$ while $E_{2n}\sim\pi^2 n^2$.
However, for the special exactly solvable case $k=0$, where the level spacing
is
constant, they are the same: $E_{\rm c}\sim 4J$ and $E_{2n}\sim 4n$.

Finally, we remark that SQES models are a generalization of the Ising limit in
quantum mechanics or quantum field theory. The Ising limit is an asymptotic
balance between a mass term and a self-interaction term: For the potential
$gx^4
-m^2x^2$ we take $g\sim m^2$ with $g$ large and get a double-well; $g$ scales
out, and we obtain the Ising limit. For SQES models there is an asymptotic
balance for large $J$ among all the coefficients of the potential.

CMB thanks the Physics Department at the Technion -- Israel Institute of
Technology and the Theoretical Physics Group at Imperial College, London, for
their hospitality, and he thanks the Lady Davis Foundation, the Fulbright
Foundation, the PPARC, and the U.S. Department of Energy for financial support.
GVD thanks the Physics Department at the University of Wales Swansea for their
hospitality and the U.S. Department of Energy for financial support. MM thanks
the SLAC theory group for their hospitality and the BSF, the ISF, and the
Technion -- VPR Fund for financial support.


\begin{references}
\bibitem{SIM} This is the simplest SQES hierarchy, just as the QES potential in
Eq.~(\ref{e1}) is the simplest of the $x^6$ QES potentials.
\end{references}
\end{document}